\documentclass[prl,showpacs,twocolumn]{revtex4}

\usepackage{epsfig}
\usepackage{color}
\usepackage{mhchem}
\usepackage{graphicx}
\usepackage{color}

\usepackage{amssymb,amsfonts,amsmath}
\usepackage{amssymb,amssymb,graphics,graphicx,amsfonts,amsmath,empheq}
\usepackage{pdfpages}

\def\d{\mathrm{d}}
\def\e{\mathrm{e}}

\begin{document}

\title{The free energy cost of reducing noise while maintaining a high sensitivity}

\author{Pablo Sartori$^1$ and Yuhai Tu$^2$.}
\affiliation{$^1$Max Planck Institute for the Physics of Complex
  Systems. Noethnitzer Strasse 38 , 01187, Dresden,
  Germany. $^2$IBM T.J. Watson Research Center, 1101 Kitchawan Road, Yorktown Heights, New York 10598, USA,.}

\begin{abstract}
Living systems need to be highly responsive, and also to keep fluctuations low. These goals are incompatible in equilibrium systems due to the Fluctuation Dissipation Theorem (FDT). Here, we show that biological sensory systems, driven far from equilibrium by free energy consumption, can reduce their intrinsic fluctuations while maintaining high responsiveness. By developing a continuum theory of the {\it E. coli} chemotaxis pathway, we demonstrate that adaptation can be understood as a non-equilibrium phase transition controlled by free energy dissipation, and it is characterized by a breaking of the FDT. We show that the maximum response at short time is enhanced by free energy dissipation. At the same time, the low frequency fluctuations and the adaptation error decrease with the free energy dissipation algebraically and exponentially, respectively.

\end{abstract}

\maketitle



Living organisms need to respond to external signals with high sensitivity, and at the same time, they also need to control their internal fluctuations in the absence of signal. In equilibrium systems, the fluctuation dissipation theorem (FDT) dictates that these two desirable properties, high sensitivity and low fluctuation, can not be satisfied simultaneously. Most sensory and regulatory functions in biology are carried out by biochemical networks that operate out of equilibrium -- metabolic energy is spent to drive the dynamics of the network \cite{qian2007phosphorylation, mehta2012energetic,niven2008energy, bennett1979dissipation}.  Thus, in principle they are not constrained by the FDT \cite{martin2001comparison}. However, how fluctuations, energy dissipation, and sensitivity are related for such systems remains not well understood. Here, we address this question by studying a negative feedback network responsible for adaptation in the bacterial chemosensory system \cite{berg1972chemotaxis,block1983adaptation,barkai1997robustness,tu2013quantitative}. 

\begin{figure}[htb]
\centerline{\includegraphics[]{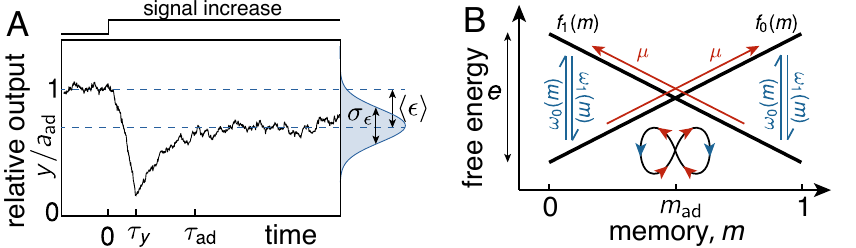}}
\caption{{\bf Noisy response of feedback adaptation.}
A) Adaptive output response to a step input signal increase at time $0$. After a sharp response in a time $\tau_y$, the output $y$ recovers back in a time $\tau_{\rm ad}$ to its adapted value $a_{\rm ad}$. The adaptation error is characterized by its average $\langle\epsilon\rangle$, as well as its variance $\sigma_\epsilon$.  B) Schematic of the feedback adaptation model. Transitions between the active and inactive memory energy landscapes, $f_1$ and $f_0$, are mediated via equilibrium activity transitions with rates, $\omega_{0}$ and $\omega_1$. An external energy input $\mu$ is used to drive the memory variable uphill in both the active and inactive states.  The result is a dissipative loop of probability flow around the adapted memory state $m_{\rm ad}$, which ensures the output to be near $a_{\rm ad}$.
\label{fig:scheme}}
\end{figure}

A typical adaptive behavior in a small system such as a single cell is shown in Fig.~\ref{fig:scheme}A \cite{koshland1982amplification}. In response to a change of the signal $S$, the output $y$ of the sensory system first changes quickly with a fast time scale $\tau_y$. After the fast response, the output slowly adapts back towards its pre-stimulus level $a_{\rm ad}$ with an adaptation time $\tau_{\rm ad}\gg \tau_y$. The new steady state (adapted) output may differ from the pre-stimulus value, and the difference is quantified by the adaptation error $\epsilon$. In our previous work \cite{lan2012energy}, we showed that the negative feedback network responsible for adaptation operates out of equilibrium with a finite free energy dissipation rate $\dot{W}$. The average adaptation error $\langle\epsilon\rangle$ was found to decrease exponentially with $\dot{W}\tau_{\rm ad}$. However, how the variance $\sigma^2_\epsilon$ of the error behaves in an adaptive system still remains unknown. This is an important question as adaptive feedback systems are intrinsically noisy due to the slow adaptation dynamics \cite{sartori2011noise}.

In the linear response regime, the output response of a system to an input signal $S(t)$ is given by $R(t) =R(0)+\int_0^t\chi (t-t')S(t')\d t'$, where $\chi$ is the response function.  For equilibrium systems, under the general assumption that response and signal are conjugate variables, the FDT establishes that $\chi (t)=-\beta \partial_t C_R (t)\Theta(t)$, where $C_R(t)\equiv \langle R(t)R(0)\rangle-\langle R\rangle ^2$ is the auto-correlation function, $\Theta(t)$ is the Heaviside function, and $\beta=(k_BT)^{-1}$ is the inverse thermal energy set to unity hereafter. For a small step stimulus $S(t)=S_0\Theta(t)$, integration of the FDT leads to a relation between the response and its correlation: $R(t)=R(0)-S_0 (C_R(t)-C_R(0))$. Since for equilibrium systems $C_R(t)$ is a monotonically decreasing function of time \cite{VanKampen}, the response $R(t)$ is also monotonic in time, and thus no adaptation dynamics is possible. Furthermore, the long time response $\Delta R\equiv R(t=\infty)-R(0)$ is linearly proportional to the variance $\sigma_R^2= C_R(0)$, i.e., $\Delta R=S_0\sigma_R^2$. 

In this paper, we show that in a non-equilibrium adaptive system both the average adaptation error $\langle\epsilon\rangle$ (analogous to $\Delta R$) and its variance $\sigma_\epsilon^2$ (analogous to $\sigma_R^2$) are suppressed by the free energy dissipation of the system but in different ways, which results to a nonlinear (logarithmic) relationship between them.
More importantly, violation of the FDT allows suppression of noise without compromising the strength of the fast response.

{\em The continuous model of feed-back adaptation.}  We start by introducing a discrete adaptation model motivated by the {\it E. coli} chemotaxis pathway. The system is characterized by its binary receptor activity $A=0,1$, which determines the output $y$; and an internal control variable $M= 0,1,\ldots,N$, that corresponds to the chemoreceptor's methylation level in {\it E. coli} chemotaxis \cite{tu2013quantitative}. For a given external input signal $S$, the free energy of the system can be written as:
\begin{align}
F_A(M,S)=-(A-1/2)[(M-M_{\rm r})E-(S-S_{\rm r})],
\label{eq:energy}
\end{align}
where $S_{\rm r}$ is a reference signal at a methylation level $M_{\rm r}$, and $E$ sets the methylation energy scale. For {\it E. coli} chemotaxis, the signal $S$ depends on the ligand attractant concentration logarithmically \cite{kalinin2009logarithmic}.

The dynamics of the system is characterized by the transitions between the $2\times(N+1)$ states in the $A\times M$ phase space. The receptor activity switches at a time scale $\tau_a$, which is much shorter than the adaptation time scale $\tau_{\rm ad}$ at which the internal variable $M$ is controlled.  The receptor activity $A$ determines the output $y$ of the signaling pathway. In the case of {\it E. coli} chemotaxis, this is carried out by the phosphorylation and dephosphorylation reactions of the response regulator CheY with an intermediate time scale $\tau_y$: $\tau_{\rm ad}\gg\tau_y\gg\tau_a$. To account for this, we relate $A$ and $y$ by $y(t)=\tau_y^{-1} \int_{-\infty}^t\e^{(t'-t)/\tau_y}A(t')\d t'$, which averages the fast binary activity over the time scale $\tau_y$.

According to Eq.~(\ref{eq:energy}), a larger signal $S$ favors the inactive state $A=0$. Thus, an increase in $S$ quickly reduces the system's average activity, at time scale $\sim\tau_a$, and output, at time scale $\sim\tau_y$, as represented in Fig.~\ref{fig:scheme}A. After this sudden initial response, the system slowly adapts by adjusting its internal variable $M$ to balance the effect of the increased signal. Due to its slow time scale, $M$ effectively serves as a memory of the system. This adaptation process restores activity and output to a level near its pre-stimulus value $\langle A\rangle=\langle y\rangle\approx a_{\rm ad}$.  Although highly precise, this adaptation process is imperfect, and its inaccuracies are quantified by the adaptation error $\epsilon$, which we define as
\begin{align}
\epsilon=\frac{y-a_{\rm ad}}{a_{\rm ad}}\quad.
\end{align}
For {\it E. coli} chemotaxis, the adaptive machinery consists of  chemical reactions that increase $M$ in the inactive state and decrease it in the active state. Note from Eq.~(\ref{eq:energy}) that such regulatory reactions are energetically unfavorable, and thus require a chemical driving force $\mu$, see Fig.~1B. 

To gain analytical insights about dynamics and energetics of adaptation, we consider the limit where $N\to\infty$ and $m=M/N\in[0,1]$ becomes a continuous variable  \cite{gardiner1986handbook}. Note that free energy and bare rates need to be rescaled for the continuum limit to converge (see Supplementary Information, SI, for details). Proceeding in this way we obtain two coupled Fokker-Planck equations that describe the chemotaxis pathway dynamics:
\begin{align}
\partial_t p_1 &= p_{0}\omega_{0} - p_{1}\omega_{1}-\partial_mJ_{1}\nonumber\\
\partial_t p_0 &= p_{1}\omega_{1} - p_{0}\omega_{0}-\partial_mJ_{0},
\label{eq:fpe}
\end{align}
where $p_1(m,t)$ and $p_0(m,t)$ are the joint probabilities for the active and inactive states with a given $m$ respectively. 
The probability currents are given by
\begin{align}
J_{A}= D_A \left((-\partial_m f_{A}-\mu)p_A-\partial_m p_A\right),\;\;\;\; A=0,1,
\end{align}
where $f_{A}(m)=-(A-1/2)[(m-m_r)e-(S-S_r)]$ is the continuum limit of Eq.~(\ref{eq:energy}) characterized by the rescaled energy parameter $e=NE$. The fast transition rates between the active and inactive states, $\omega_0$ and $\omega_1$, satisfy detailed balance $\omega_0/\omega_1=\exp(f_0-f_1)$. The diffusion-like constants $ D_1$ and $ D_0$ set the time-scale of $m$ changes for active and inactive states, and thus the adaptation time goes as $\tau_{\rm ad}\sim D_A^{-1}$, see SI for details. Our model is analogous to that of an isothermal ratchet \cite{parmeggiani1999energy}, where a chemical driving fuels directed motion. Whereas in ratchets $\mu$ drives directed motion, here it  fuels currents up the energy landscapes $f_0$ and $f_1$ to achieve adaptation.

In the absence of external driving, i.e. $\mu=0$, the system relaxes to a state of thermal equilibrium with no phase-space fluxes $J_0=J_1=0$. In this regime  adaptation is impossible. The chemical driving $\mu>0$ breaks detailed balance and creates currents that increase $m$ in the active state and decrease it in the inactive state. For large enough $\mu$, the memory variable $m$ can be stabilized (trapped) in a cycle around its adapted state $m_{\rm ad}$, which ensures $\langle y\rangle\approx a_{\rm ad}$ as illustrated in Fig.~1B. The free energy dissipation rate $\dot{W}$ can be computed $\dot{W}\approx C{|\mu|}/{\tau_{\rm ad}}$ with $C$ a system specific constant set to unity by our parameter choice, see SI. In the following, we will use the chemical driving $\mu\approx\tau_{\rm ad}\dot{W}$ to characterize the system's energy dissipation.




\begin{figure}[!htb]
\centerline{\includegraphics[]{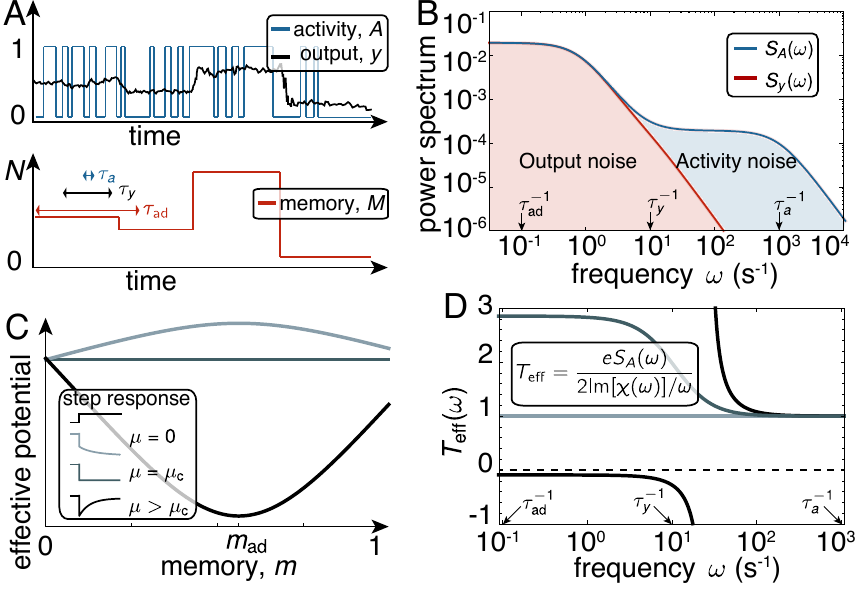}}
\caption{{\bf Adaptation as a non-equilibrium  transition} A) Schematic time traces of the binary activity $A$ (blue), the output $y$ (black), and the memory $M$ (red) in steady state. The slow $M$ variations induce large fluctuations in the output $y$, while the fast $A$ switching for a fixed $M$ only produces small fluctuations in $y$. B) Power spectra of the activity $S_A$ and output $S_y$. The output noise is filtered (reduced) in the high frequency range $\tau_y^{-1}<\omega<\tau^{-1}_a$; but it remains unfiltered in the range $\tau_{\rm ad}^{-1}<\omega<\tau^{-1}_y$ . C) Effective memory potential in Eq.~\ref{eq:neqpotential} for three values of the chemical driving $\mu$ (due to the choice $D_1=D_0$ taken here, $m_{\rm ad}=m_*$). At equilibrium, $\mu=0$, the adapted memory state $m_{\rm ad}$ is unstable. At the value $\mu=\mu_{\rm c}$ the system becomes critical. In the region $\mu>\mu_{\rm c}$ the  adapted state $m_{\rm ad}$ is stable. Inset: Activity response to step signal increase for corresponding values of $\mu$.  D) Effective temperature  $T_{\rm eff}$ for three different values of the chemical driving $\mu$. After the onset of adaptation a region with ``negative friction''  develops, at the end of which the effective temperature diverges. Values of $\mu$ from lighter to darker blue are $\mu=0$, $\mu=0.65\mu_{\rm c}$, and $\mu=20\mu_{\rm c}$ (the same as in panel C). The other parameters are  from \cite{Tu30092008}, see SI.} \label{fig:fdt}
\end{figure}

The dynamics of $A$, $y$, and $m$ are illustrated in Fig. 2A.  
In the power spectra of Fig. 2B, the high frequency fluctuation of $y$ is suppressed from that of $A$ by time-averaging. However, the low frequency fluctuations, which are caused by the slow fluctuations of $m$, are not affected. These low frequency noise can be suppressed by free energy dissipation, as we show later in this paper.

{\em Adaptation as a non-equilibrium phase transition.}
Given the separation of time scales $\tau_a\ll\tau_{\rm ad}$, we can solve Eqs.~(\ref{eq:fpe}) by using the adiabatic approximation \cite{VanKampen,Tu30092008}:
$p_1(m)=a(m)p(m)$, and $p_0(m)=(1-a(m))p(m)$,
with $a(m)=(1+\e^{f_1(m)-f_0(m)})^{-1}$ the average equilibrated activity for a fixed value of $m$. The distribution of $m$ can be written as $p(m) = \e^{-h(m,S)}/Z$ with $h$ the effective potential and $Z$ a normalization constant. We have determined the effective potential $h$ analytically (see SI for detailed derivation):
\begin{align}
h(m,S)=\frac{\mu}{\mu_c}\ln[ D_0 e^{-(m-m_*)e/2}
+ D_1 e^{(m-m_*)e/2}]\nonumber\\ -\ln[ e^{-(m-m_*)e/2}+ e^{(m-m_*)e/2}]\quad,
\label{eq:neqpotential}
\end{align}
where we have defined the critical chemical driving as $\mu_{\rm c}=e/2$, and $m_*=m_{\rm r}+(S-S_{\rm r})/e$. 

The analytical form of the effective potential is one of the main results of this paper. The effect of  energy dissipation and the onset of adaptation can be understood intuitively with $h(m,S)$, which contains two terms with similar shapes, see Fig.~\ref{fig:fdt}C. The first term (proportional to $\mu/\mu_{\rm c}$)
in the right hand side of Eq.~(\ref{eq:neqpotential}) comes from chemical driving (non-equilibrium effect) and  has a stable free energy minimum. The second term is the equilibrium potential in the absence of driving, and has a maximum at $m_*$. At equilibrium the only critical point $m_*$ is unstable, so the system tends to go to the boundaries without adapting.  
As $\mu$ increases the first part of the potential starts to dominate. For $\mu>\mu_{\rm c}$, the system develops a stable fixed point at $m_{\rm ad}$ away from the boundaries, indicating the onset of adaptive behaviors \cite{Allahverdyan2013}. As $\mu$ increases this fixed point becomes increasingly stable, and adaptation accuracy increases.   
The transition of a feedback system to adaptation can thus be loosely understood as a continuous phase transition (see SI for details). Since the control parameter is the free energy dissipation, the transition to adaptation occurs far from equilibrium and a breaking of FDT is to be expected.

{\em Breakdown of Fluctuation Dissipation Theorem.}
In our feedback model, the observable conjugate to the signal is  $e A=-\partial_Sf_A$, but see \cite{buijsman2012efficient, de2013unraveling, sartori2014thermodynamic} for cases where this is not true. The FDT would lead to $\chi(t)=e\partial_tC_{A}(t)$, where $\chi$ is the activity response function and $C_A$ is the monotonic correlation function. In an adaptive system the integral of $\chi$, which is just the response to a step stimulus, is non-monotonic, therefore FDT is broken.

To quantify the departure from equilibrium, we define an effective temperature $T_{\rm eff}$ using the formulation of the FDT in frequency space \cite{cugliandolo1997energy,martin2001comparison}, see inset in Fig.~\ref{fig:fdt}D. The frequency-dependence of $T_{\rm eff}$ for $\mu>0$ implies a breakdown of FDT. As shown in Fig.~\ref{fig:fdt}D, while for any value $\mu\neq0$ we have $T_{\rm eff}\neq1$, after the transition to the adaptive regime $\mu\ge\mu_{\rm c}$ a divergence occurs. This corresponds to the appearance of a frequency region where ${\rm Im}[\chi(\omega)]<0$. A negative effective viscosity indicates the dominance of the active effects that drive a net current to flow against the gradients of the equilibrium energy landscape $f_A$, something also observed in other biological systems such as collections of motors \cite{julicher1997spontaneous} or the inner ear hair bundle \cite{martin2001comparison}. The breakdown of FDT means that there is no {\it a priori} connection among fluctuations $\sigma_\epsilon^2$, chemical driving $\mu$, and long-time response $\langle\epsilon\rangle$. In the following we derive relations linking these three quantities in the adaptive feedback system studied here.

{\em The free energy cost of suppressing fluctuations.} As evident from the effective potential, increasing the chemical driving $\mu$ stabilizes the adapted state. In the limit $\mu\rightarrow \infty$, the system thus goes to its perfectly adapted state with average output $a_{\rm ad}=D _0/( D_0+ D_1)$. For finite $\mu$, the output differs from $a_{\rm ad}$, which can be characterized by the average error $\langle\epsilon\rangle$ and its variance $\sigma_\epsilon^2$.

\begin{figure}[!htb]
\begin{center}
\includegraphics[]{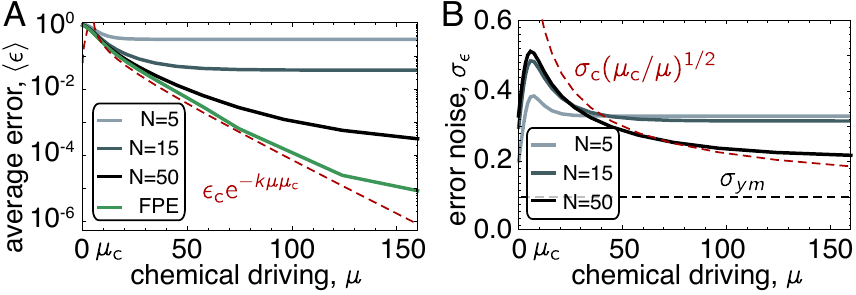}
\caption{{\bf Free energy cost of reducing error and noise. } A). Dependence of average error with chemical driving for several system sizes. The decay is  exponential, in agreement with the infinite size limit (dashed red). Saturation of the decay for finite $N$ is due to finite size effects. B) Adaptation noise as a function of chemical driving for several system sizes, together with the analytical estimate in dashed red. At very large driving the noise saturates to its minimum $\sigma_{ ym}$ dictated by the intrinsic activity fluctuations. Note that at the critical driving $\mu_{\rm c}$ the analytical estimate diverges. This divergence is smoothed for finite $N$. \label{fig:esaesf}}
\end{center}
\end{figure}

The average adaptation error is $\langle \epsilon\rangle =( \langle y\rangle-a_{\rm ad})/a_{\rm ad}$. Summing and integrating Eqs.~(\ref{eq:fpe}), we have
\begin{align}
\langle \epsilon \rangle= \frac{ D_1 p_1(1)+ D_0p_0(1)}{ D_0(e/2-\mu)}- \frac{ D_1 p_1(0)+ D_0p_0(0)}{ D_0(e/2-\mu)}\quad. \label{eq:errorgen}
\end{align}
Thus to obtain the adaptation error we only need to evaluate the probability at the boundaries. In the limit of $\mu\gg\mu_c$, we have:
\begin{align}
\langle\epsilon\rangle\approx\epsilon_{\rm c}\e^{-k\mu\mu_{\rm c}}\quad,\label{eq:esa}
\end{align}
where $k$ and $ \epsilon_{\rm c}$ are constants with only weak dependence on $\mu$ (see SI for derivation). This shows explicitly that the adaptation error goes down exponentially with energy dissipation, as found numerically in our previous work for the discrete model \cite{lan2012energy}. Here, we show this relationship analytically in the continuum limit. Fig.~\ref{fig:esaesf}A shows results from both the continuum and discrete model.


Besides stabilizing the adapted state, Eq.~(\ref{eq:neqpotential}) shows that increasing $\mu$ also reduces the $m-$fluctuations by making the effective potential sharper. The reduction in these fluctuations implies a decrease in the variance of the error $\sigma_\epsilon^2$.
Taking into account the separation of time scales $\tau_a\ll\tau_y\ll\tau_{\rm ad}$, the variance of  the output $y$ can be approximated as the sum of two variances $\sigma_{ym}^2$ and $\sigma_a^2$. They respectively correspond to variation of $y$ at time scale $\sim \tau_y$ around its average $a(m)$ for a fixed $m$, and the variation of $a(m)$ due to variation of $m$ at the adaptation time $\sim\tau_{\rm ad}$. We thus have
\begin{align}
\sigma^2_\epsilon \approx (\sigma_{ym}^2+\sigma_a^2)/a_{\rm ad}^2\quad.\label{eq:esfsistand}
\end{align}
The variance $\sigma^2_{ym}$ of $y$ is caused by the fast fluctuations of the binary variable $A$ at timescale $\sim\tau_a$ averaged over the output timescale $\tau_y\gg \tau_a$ (see SI for derivation):
\begin{align}
\sigma^2_{ym}= (a_{\rm ad} -a_{\rm ad}^2) \tau_a/(\tau_y+\tau_a)\quad,\nonumber
\end{align}
which clearly shows that $\sigma^2_{ym}\propto \tau_a/\tau_y$ is reduced by time-averaging.

The variance $\sigma_a^2=\langle a\rangle^2-\langle a^2\rangle$, where $\langle a^n\rangle=\int_0^1 a^n(m)p(m)\d m$ for $n=1,2$, is caused by the slow variation of $m$. To obtain an analytical expression for $\sigma_a^2$ we approximate $p(m)$ by a Gaussian, valid for $\mu\gg\mu_{\rm c}$. This results in $\sigma_a^2\approx (\partial_m{a}_{\rm ad})^2\sigma_m^2/a_{\rm ad}^2$.  The variance $\sigma_m^2=\langle m\rangle^2-\langle m^2\rangle$ within the same Gaussian approximation  of $p(m)$ is  given by $\sigma_m^2\approx (\mu\mu_{\rm c})^{-1}$. Defining now a characteristic variance as $\sigma_{\rm c}^2=4(1-a_{\rm ad})^2a_{\rm ad}^2$, we finally have:
\begin{align}
\sigma_a^2\approx \sigma_{\rm c}^2{\mu_{\rm c}}/{\mu}\quad,\label{eq:esf}
\end{align}
which vanishes when $\mu\to\infty$. This is a main result of the paper, which shows that energy dissipation is used to reduce error noise by suppressing slow activity fluctuations. Figure 3B compares this expression to several discrete models with increasing $N$.



\begin{figure}[!htb]
\begin{center}
\includegraphics[]{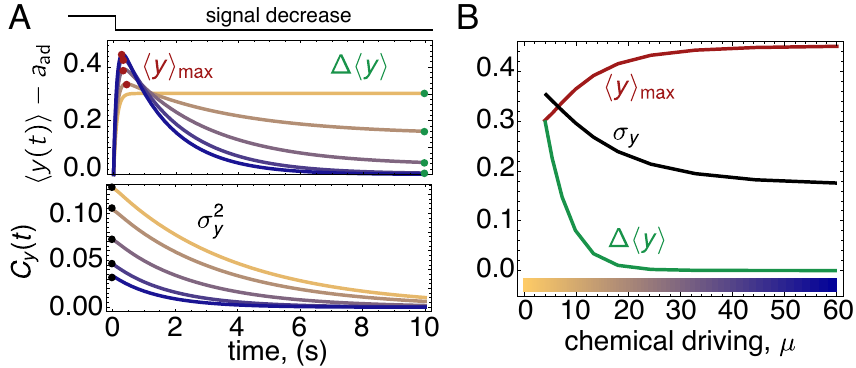}
\caption{ {\bf Response and correlations in systems out of equilibrium.}   A) (top panel) Average output response to a signal decrease for several values of the chemical driving beyond $\mu_{\rm c}$, see the color-code for $\mu$ in panel B. As the chemical driving $\mu$ increases, the maximal transient response $\langle y\rangle_{\rm max}$ increases, but the long time response $\Delta\langle y\rangle=a_{\rm ad}\langle \epsilon\rangle$ decreases. (bottom panel) The correlation function also decreases as the system is driven further away from equilibrium. B) The dependence of $\langle y\rangle_{\rm max}$, $\Delta\langle y\rangle$, and $\langle y\rangle_{\rm max}$ on the chemical driving $\mu$. The long-time response (adaptation error) $\Delta\langle y\rangle$ decreases quickly with $\mu$. The decrease of the output fluctuation (noise) $\sigma_y$ with $\mu$ is more gradual, and controls the increase in the maximal response $\langle y\rangle_{\rm max}$ for large $\mu$. In this figure $N=15$, and $S=S_{\rm r}$.  \label{fig:takehome}}
\end{center}
\end{figure}

\emph{Discussion.} {Biochemical networks are non-equilibrium systems fueled by free energy dissipation to achieve their biological functions. Energy dissipation liberates the networks from constraints such as the Fluctuation Dissipation Theorem and Detailed Balance. Here, we show in a negative feedback network that the long-time output response $\Delta\langle y\rangle=a_{\rm ad}\langle\epsilon\rangle$ decreases with the free energy dissipation $\mu\approx\tau_{\rm ad}\dot{W}$ exponentially, and its fluctuation $\sigma_y^2=a_{\rm ad}^2\sigma_\epsilon^2$ decreases as $\mu^{-1}$. Both these effects, especially the slower decay of $\sigma_y^2$ with $\mu$, contribute to enhance the short time response $\langle y\rangle_{\rm max}$, see  Fig.~\ref{fig:takehome}. 

Even though FDT is broken in the adaptive system studied here, fluctuations and long-time response of the output are linked via a non-linear relation: $\sigma_y^2\approx d\mu_{\rm c}/\log(y_{\rm c}/\Delta\langle y\rangle)+\sigma_{ ym}^2$, where $d=k\sigma_{\rm c}^2$ and $y_{\rm c}=a_{\rm ad}\langle\epsilon\rangle$. Unlike the linear non-equilibrium FDT derived by a change of observables \cite{bedeaux1971linear,prost2009generalized,seifert2010fluctuation}, our non-linear relation links observables that are conjugate at equilibrium, making it particularly appealing. Another approach is taken in \cite{barato2015thermodynamic}, where near equilibrium linear response is used to show that the dispersion of variables can be reduced by dissipation. Adaptation however is a far from equilibrium phenomenon which requires a critical finite amount of free energy dissipation. As a result, the energy scale is set by the intrinsic energy $\mu_{\rm c}$  instead of the thermal energy $k_{\rm B} T$ in \cite{barato2015thermodynamic}. It remains a challenging question whether these approaches can be combined to obtain a general relationship among response, fluctuation, and energy dissipation for systems far from equilibrium.

\begin{acknowledgments}
This work is partly supported by a NIH grant (R01GM081747 to YT). We thank Leo Granger and Jordan Horowitz for a critical reading of this manuscript.
\end{acknowledgments}

\clearpage

\section{ Supplementary Material}
\subsection{Continuum limit of feedback adaptation.} The out-of-equilibrium dynamics of the discrete model in its phase space $A\times M \in \{[0,1],[0,N]\}$ are governed by six sets of rates. The rates $\omega_0(M)$ govern the transitions from the inactive $A=0$ states to the active $A=1$ states, and the rates  $\omega_1(M)$ the reciprocal inactivation transitions. The memory of active states is increased with a rate $k^+_1(M)$ and decreased with a rate $k^-_1(M)$. For inactive states, the memory can increase with a rate $k^+_0(M)$ and decrease with $k^-_0(M)$.

Given the free energy from Eq.~\ref{eq:energy}, we have that the rates of passive activity transitions satisfy detailed balance,
\begin{align}
\frac{\omega_1(M)}{\omega_0(M)}=\e^{F_1(M)-F_0(M)}=\e^{-[(M-M_{\rm r})E-(S-S_{\rm r})]}\quad.\nonumber
\end{align}
The memory transitions are driven out of equilibrium, and in general we have
\begin{align}
\frac{k^+_0}{k^-_0}&=\e^{F_0(M)-F_0(M+1)+G}={\e^{-E/2+G}}\nonumber\\
\frac{k^+_1}{k^-_1}&=\e^{F_1(M)-F_1(M+1)+G}=\e^{E/2-G},\nonumber
\end{align}
where the ratios are independent of $M$, and $G$ is the free energy input in the reactions which keeps them out of equilibrium. When there is no free energy input the system satisfies detailed balance and its dynamic are simple equilibrium relaxation. For large values of $G$ the memory only increases for inactive states and decreases for active ones, the chemotaxis limit \cite{lan2012energy}.

The dynamics of this system are governed by the master equation. For the bulk states it is best written as two coupled equations
\begin{align}
\partial_t{ P_1(M)} &= P_0(M)\omega_{0}(M)+P_1(M-1)k^+_1\nonumber\\
& + P_1(M+1)k^-_1- P_1(M) [\omega_1(M) + k^+_1 + k^-_1]\quad,\nonumber\\
\partial_t{ P_0(M)} &= P_0(M)\omega_{0}(M)+P_1(M-1)k^+_1\nonumber\\
& + P_1(M+1)k^-_1- P_1(M) [\omega_1(M) + k^+_1 + k^-_1]\quad.\nonumber
\end{align}
For the upper boundary $M=N$ we have
\begin{align}
\partial_t{ P_A(N)} &= P_{1-A}(N)\omega_{1-A}(N)+P_A(N-1)k^+_A\nonumber\\
& - P_A(N) [\omega_A(N)  + k^-_A]\quad,\nonumber
\end{align}
and analogously for the lower boundary $M=0$, we have
\begin{align}
\partial_t{ P_A(0)} &= P_{1-A}(0)\omega_{1-A}(N)+P_A(1)k^-_A\nonumber\\
& - P_A(0) [\omega_A(0)  + k^+_A]\quad.\nonumber
\end{align}
A general solution of the master equation can be obtained using standard linear algebra \cite{VanKampen}.

To obtain the continuum theory we perform an expansion in the number of memory states $N$, and then take the limit $N\to\infty$. This changes the discrete variable $M\in[0,N]$ to the continuous $m\in[0,1]$. The probability density is defined by $p_A(m)=P(A,M)/N$, where $m=M/N$. Furthermore, for the continuum limit to exist, the parameters $\mu=N G$, $e=NE$ and $m_{\rm r}=M_{\rm r}/N$ have to be kept constant as $N\to\infty$. Using this, the continuous free energy is simply given by
\begin{align}
f_A(m)=-(A-1/2)[(m-m_{\rm r})e-(S-S_{\rm r})]\quad,\nonumber
\end{align}
which through detailed balance defines the continuous activation and inactivation rates $\omega_0(m)$ and $\omega_1(m)$ up to a time scale. Dividing now the bulk master equations by $N$ and expanding to second order in $1/N$ results in two chemically coupled Fokker-Planck equations
\begin{align}
\partial_tp_1 &= p_0\omega_0 - p_1\omega_1-H_1 \partial_mp_1 +  D_1\partial^2_mp_1\nonumber\\
\partial_tp_0 &= p_1\omega_1 - p_0\omega_0-H_0 \partial_mp_0 +  D_0\partial^2_mp_0\quad,\label{eq:fpesi}
\end{align}
where the drift and diffusion coefficients are given by
\begin{align}
H_A=\lim_{N\to\infty}\frac{k_A^+-k_{A}^-}{N}\;\;;\;\; D_A=\lim_{N\to\infty}\frac{k_A^++k_{A}^-}{2N^2}\quad.\nonumber
\end{align}
To calculate the first limit, we can use that
\begin{align}
\frac{k_A^+-k_{A}^-}{N}=2 N\frac{k_A^++k_{A}^-}{2N^2}\frac{1-k^-_A/k^+_A}{1+k^-_A/k^+_A}\quad,\nonumber
\end{align}
and expand in $1/N$ the ratio of rates, which results in $H_{0}=-(e/2-\mu) D_0$ and $H_{1}=(e/2- \mu) D_1$. Note that $H_0= D_0(-\partial_mf_0+\mu)$ and $H_1= D_1(-\partial_mf_1-\mu)$, where the difference in sign in front of the chemical driving is due to the fact that memory transitions  are driven in opposing directions for active and inactive states.

Finally, proceeding in the same way on the discrete boundary equations gives at $m=1$ and $m=0$ the boundary condition
\begin{align}
0 &= H_Ap_A -  D_A\partial_mp_A\quad,\nonumber
\end{align}
which is nothing but a no-flux condition. A verification of the convergence of the discrete to the continuous solution on the steady state appears in Fig.~\ref{fig:si1} A, where the discrete model was solved numerically for increasing values of $N$.

\begin{figure}[!htb]
\begin{center}
\includegraphics[]{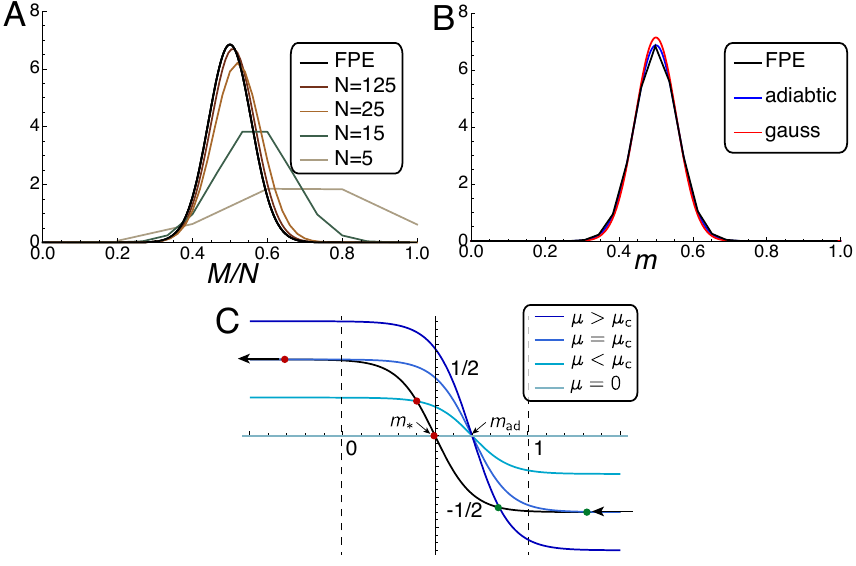}
\caption{ { A)} Convergence of the steady state solution of the ME $P(M)$ to the FPE solution $p(m)$. Parameters as elsewhere in the text, with $\mu=20\mu_{\rm c}$ and $S=S_{\rm r}$. { B)} Numerical solution of the FPE (black) and two analytical approximations, the Gaussian (red) and the adiabatic using the potential $h$. Parameters as in B. { C)}  In black, the equilibrium contribution to $\partial_m h$, and in shades of blue the non-equilibrium contribution for different values of $\mu$. Both in units of $e$ and as a function of $m$, with the range $[0,1]$ delimited by dashed lines. Sinc these terms have opposing sign, see Eq.~\ref{eq:dh}, extrema correspond to points where a blue line crosses the black line, and if the slope of the blue curve is higher than that of the black one they are minima. As one can see, the maximum (in red) disappears through the left boundary, while the minimum (in green) comes through the right one. This feature makes the transition second order.}  \label{fig:si1}
\end{center}
\end{figure}

\subsection{Self-consistent choice of diffusion constants}

The constants $ D_0$ and $ D_1$ with units of frequency are analogous to the diffusion constant of a random walk. They define the characteristic time scale of the $m$-dynamics, as well as the value of the adapted activity.  In fact, the adapted activity for the discrete model is reached when $G\to\infty$ and $k_1^+=0=k_0^-$, which results in $a_{\rm ad}=k_0^+/(k_0^++k_1^-)$ \cite{lan2012energy}. Using the expressions for $ D_0$ and $ D_1$, this gives
\begin{align}
a_{\rm ad}=\frac{ D_0}{ D_0+ D_1}\quad.\label{eq:acadap}
\end{align}
The time scale of adaptation $\tau_{\rm ad}$ defines the rate at which the memory relaxes, which in the discrete case is given by  $k_0^++k_1^-$. In the continuum limit however the relaxation rate of the memory is given by $(e-\mu)( D_0+ D_1)$. Since accurate adaptation is reached in the regime $\mu\to\infty$, which should not affect the adaptation time, we choose the diffusion-like constants as
\begin{align}
D_A=\frac{\bar{D}_A}{\tau_{\rm ad}(e+\mu)}\nonumber\quad,
\end{align}
with the dimensionless constants $\bar{D}_A$ of order one chosen such that Eq.~\ref{eq:acadap} is satisfied. 

Note that this choice of diffusion constants limits the choice of rates in the discrete model, of which we have so far only specified the ratios. One choice of time scale compatible with the $ D_A$ above is
\begin{align}
k_A^++k_A^-=\frac{2N}{\tau_{\rm ad}}\frac{1+\e^{E/2+G}}{\e^{E/2+G}-1}\quad,\nonumber
\end{align}
as can be verified by taking the limit $N\to\infty$ which defines $ D_A$. For this choice, we have that in the irreversible limit the ``diffusion'' term of the master equation for finite $N$ is
\begin{align}
\lim_{G\to\infty}k_A^++k_A^-=\frac{2N}{\tau_{\rm ad}}\quad,\nonumber
\end{align}
ultimately determined by the adaptation time. The same is true for the ``drift'' of active and inactive memory states, which are given by
\begin{align}
\lim_{G\to\infty}k_0^+-k_0^-=\frac{2N}{\tau_{\rm ad}}\quad{\rm and}\quad\lim_{G\to\infty}k_1^+-k_1^-=-\frac{2N}{\tau_{\rm ad}}\nonumber\;.
\end{align}

These choices of time scales, which together with the ratio of the rates uniquely defines all rates, ensures that the adaptation time takes  a value of order $\tau_{\rm ad}$ in the discrete and continuum models alike. Indeed, as can be seen in Fig.~\ref{fig:si1}~D, the response of the activity for a step change in ligand in the adaptive regimes is essentially the same for the continuous and discrete cases.

\subsection{Perturbative expansion for fast activity transitions}

Even at the steady state, Eqs.~\ref{eq:fpesi} are hard to solve analytically without any further assumption. In most sensory adaptive systems however there is a clear separation between the adaptation time $\tau_{\rm ad}$ and the activation time $\tau_a\ll\tau_{\rm ad}$, which we can use to obtain approximate solutions of $p_A(m)$. Consider that  $\omega_{A}= \bar{\omega}_{A}/\tau_a$, where from now on bars will denote dimensionless quantities, and $\bar{\omega}$ are of order unity. We can then define the parameter $\delta=\tau_a/\tau_{\rm ad}$, and write the dimensionless steady state equations
\begin{align}
 p_{0}\bar{\omega}_{0} - p_{1}\bar{\omega}_{1}-\frac{\delta}{e+\mu}\partial_m\bar{J}_{1}&=0\nonumber\\
 p_{1}\bar{\omega}_{1} - p_{0}\bar{\omega}_{0}-\frac{\delta}{e+\mu}\partial_m\bar{J}_{0}&=0\quad,\label{eq:pert1}
\end{align}
where the dimensionless fluxes are defined using $\bar{D}_A$. In addition to these equations, because of total flux conservation, we have that $\partial_m(\bar{J}_0+\bar{J}_1)=0$, which together with the boundary conditions gives
\begin{align}
\bar{J}_0+\bar{J}_1=0\label{eq:pert2}\quad.
\end{align}
Since typically in adaptive sensory systems $\delta\ll1$, it is natural to expand the steady state probabilities as
\begin{align}
p_A=p_A^{(0)}+\delta p_A^{(1)}+\delta^2p_A^{(2)}+\ldots
\end{align}
We now provide the equations  for the first two terms of this expansion.

\textit{Zeroth order.} We begin by using the definition of conditional probability, which gives $p^{(0)}_1=a^{(0)}p^{(0)}$ and $p^{(0)}_0=(1-a^{(0)})p^{(0)}$. Here $a^{(0)}$ is the conditional activity distribution to zeroth order, which we call $a(m)$ in the main text; and $p^{(0)}$ the marginal memory distribution to zeroth order, noted in the main text as $p(m)$. To zeroth order, Eq.~\ref{eq:pert1} establishes that
\begin{align}
\bar{\omega}_0p_0^{(0)}=\bar{\omega}_1p_1^{(0)}\quad,\nonumber
\end{align}
which is a no-flow condition for the activity transitions. From it, we can derive that $a^{(0)}=\omega_0/(\omega_1+\omega_0)$. Inserting this in Eq.~\ref{eq:pert2} gives to lowest order
\begin{align}
(\bar{H}_1a^{(0)}&+\bar{H}_0(1-a^{(0)}))p^{(0)}\nonumber\\
&-\partial_m[(\bar{D}_1a^{(0)}+\bar{D}_0(1-a^{(0)}))p^{(0)}]=0\nonumber\quad,\label{eq:fpem}
\end{align}
This equation has a solution of the form $p^{(0)}=\exp(-h)/Z$, where $Z$ is a normalization constant and $h$ is an effective non-equilibrium potential. This potential can be integrated analytically, which results in
\begin{align}
h(m,S)&=\left(\frac{m e}{2}+\log\left[ \frac{1}{1+\omega_0/\omega_1}\right]\right)\nonumber\\
&-\frac{\mu}{e/2}\left(\frac{me}{2}+\log\left[\frac{1}{ D_0/ D_1+\omega_0/\omega_1}\right]\right)\;\;.
\end{align}
The validity of this solution can be verified by substitution in Eq.~\ref{eq:fpem}. 

\textit{First order}. The correction of first order in $\delta$ is $p_A^{(1)}$. Since we preserve the normalization of the zeroth order term, the first order term has as normalization condition $\int_0^1 (p^{(1)}_0(m)+p^{(1)}_1(m))\d m=0$. From Eq.~\ref{eq:pert1}, the first order term directly gives
\begin{align}
\bar{\omega}_0p_0^{(1)}-\bar{\omega}_1p_1^{(1)}-\frac{\partial_m\bar{J}_1^{(0)}}{e+\mu}&=0\quad,\nonumber
\end{align}
which shows that to this order part of the $m$-flux is deviated as  probability currents on the activity transitions. The global flux balance in Eq.~\ref{eq:pert2} becomes
\begin{align}
\bar{H}_1p_1^{(1)}&+\bar{H}_0p_0^{(1)}-\partial_m[\bar{D}_1p_1^{(1)}+\bar{D}_0p_0^{(0)}]=0\nonumber\quad.
\end{align}
Unfortunately these two equations can not be solved analytically. In the following, analytical approximations are made using only the zeroth order contribution to the probability.

\subsection{Characterizing the adaptive transition}
To characterize the onset of adaptation, we study the extrema of the non-equilibrium potential. The condition for a point $m$ to be an extrema is $\partial_mh=0$,  where we have
\begin{align}
\partial_mh=\frac{e}{2}-\frac{e}{1+\omega_1/\omega_0}-\frac{\mu}{e/2}\left(\frac {e}{2}-\frac{e}{1+ D_0\omega_1/\omega_0 D_1}\right)\;\;.\label{eq:dh}
\end{align}
The second derivative of $h$ characterizes whether the point is a maximum or a minimum.

At equilibrium we have $\mu=0$, and the only extrema occurs at the value
\begin{align}
m_{\rm *}=m_{\rm r}+(S-S_0)/e\nonumber\quad.
\end{align}
Note that, for $m_{\rm *}$ to be an extrema it must fall in the range $[0,1]$, something which we assume from now on. At $m_{\rm *}$  we have that $a(m_{\rm *})=1/2$, which corresponds to the maximal sensitivity of the activity to changes in the signal. It is easy to show, however, that for $\mu=0$ this point is unstable, since $\partial_m^2h<0$. Because there is no other extrema, the memory will  accumulate at the boundaries, which shows that adaptation to a high sensitivity state is not possible in equilibrium.

When $\mu\neq0$, there can be two extrema of the potential $h$. At $\mu=0$ we still have $m_{\rm *}$, however at $\mu\to\infty$ we have
\begin{align}
m_{\rm ad}(\mu\rightarrow \infty)=m_{\rm ne}\equiv m_{\rm r}+(S-S_{\rm r})/e+\epsilon^{-1}\log( D_0/ D_1)\quad,
\end{align}
where $m_{\rm ne}$ is the $m$ value at the minimum of the nonequilibrium contribution to the effective free energy, i.e., the first term on the right hand side of Eq. (5) in the main text. At $m_{\rm ad}$ the activity is  $a(m_{\rm ad})= D_0/( D_0+ D_1)$, which deviates from the maximum sensitivity point. Note that for $m_{\rm ad}$ to be a minima it must fall in the range $[0,1]$, which determines the range of signals to which the system can adapt. For the case in which $m_{\rm ad}\in[0,1]$, there is a transition at intermediate values of $\mu$ from $m_{\rm eq}$ being an unstable point, and the probability accumulating to the boundaries; to $m_{\rm ad}$ being a stable state, and the system being adaptive. By using a graphical construction of $\partial_mh$, see Fig.~\ref{fig:si1} C, one can show that this transition is second order. We now describe the phase transition step by step.

Near equilibrium and still far from the critical point the memory accumulates at the boundaries. Which boundary is more stable depends on the value of the signal. As $\mu$ increases the maximum (which is $m_{*}$ at equilibrium) is displaced towards $m=0$, for $ D_0/ D_1>1$, or $m=1$ , for $ D_0/ D_1<1$. Eventually the maximum leaves the range $[0,1]$, and there is just one stable boundary (the opposite to the one through which the maximum left) where the memory accumulates. Right at the critical point $\mu_{\rm c}=e/2$ the energy landscape has no extrema in $m\in[-\infty,+\infty]$. For values $\mu>\mu_{\rm c}$ a minima develops asymptotically on the side opposite to the one through which the maxima left. Eventually this minima comes in the range $[0,1]$ through the boundary where the memory resides, and moves asymptotically to the value $m_{\rm ad}$, which is reached for $\mu\to\infty$. Because the memory is continuously taken from the boundary to the adapted point,  this phase transition is second order.

\subsection{Fully irreversible limit}
A useful limit to study the continuous equations is $\mu\to\infty$ to all orders in $\delta$. Note that in this limit Eq.~\ref{eq:pert2} becomes simply
\begin{align}
- D_1p_1(m)+ D_0p_0(m)=0\quad.\nonumber
\end{align}
This equation is satisfied by the following expressions
\begin{align}
p_1(m)&=\frac{\omega_0(m)}{\omega_0(m)+\omega_1(m)}\delta(m-m_{\rm ad})\nonumber\\
p_0(m)&=\frac{\omega_1(m)}{\omega_0(m)+\omega_1(m)}\delta(m-m_{\rm ad})\nonumber\quad,
\end{align}
which also satisfy the normalization condition. This expressions indicate that in the fully irreversible limit the memory is exactly fixed to its adapted value, as the diffusive terms vanish. The average activity is thus clearly $a_{\rm ad}$. In this limit the activity transitions are simply governed by the two states master equation
\begin{align}
\partial_tp_1=\omega_0p_0-\omega_1p_1\quad.\label{eq:twostates}
\end{align}

\subsection{Derivation of Eq.~\ref{eq:esa}}

To obtain an analytical expression of the adaptation error we use Eq. \ref{eq:pert2}. Using this equation and defining the average activity as $\langle A\rangle=\int_0^1p_1(m)\d m$ we get:
\begin{align}
\bar{H}_1\langle A\rangle+\bar{H}_0(1-\langle A\rangle)&=\bar{D}_1p_1(1)+\bar{D}_0p_0(1)\nonumber\\
&-(\bar{D}_1p_1(0)+\bar{D}_0p_0(0))\quad,\nonumber
\end{align}
which is valid to all orders. Note that { when $\mu\to\infty$} the average activity converges to the adapted value $a_{\rm ad}$ in Eq.~\ref{eq:acadap}, in agreement with the solution derived above on the fully irreversible limit. The adaptation error of the output  is defined as
\begin{align}
\epsilon=\frac{ y-a_{\rm ad}}{a_{\rm ad}}\quad.\nonumber
\end{align}
Since $\langle y\rangle=\langle A\rangle$, we can use the previous equation to show that $\langle\epsilon\rangle=\epsilon_1-\epsilon_0$, with
\begin{align}
\epsilon_1&=\frac{ D_1 p_1(1)+ D_0p_0(1)}{ D_0(e/2-\mu)}\nonumber\\
\epsilon_0&=\frac{ D_1 p_1(0)+ D_0p_0(0)}{ D_0(e/2-\mu)}\label{eq:errorgen}
\end{align}
Thus to obtain the error we just need to evaluate $p$ at the boundaries.

We calculate the zeroth order term, and for compactness note $a^{(0)}(m)$ as $a(m)$, as done in the main text. Although to this order we have an exact form for the potential, we can not exactly integrate $\e^{-h}$ to obtain the normalization $Z$. Deep into the adaptive phase, we can drop the equilibrium component of $h$ and use a saddle point approximation with expansion parameter $\mu/\mu_{\rm c}\gg1$ to evaluate $Z$:
\begin{align}
Z=\int_0^1\e^{-h(m,S)}\d m\approx\sqrt{\frac{2\pi}{|\partial_m^2h_{\rm ad}|}}\e^{-h_{\rm ad}}=\sqrt{\frac{2\pi}{\mu\mu_{\rm c}}}\e^{-h_{\rm ad}}\;,\nonumber
\end{align}
where the subindex indicates that $h$ and its derivative are to be evaluated at $m_{\rm ad}$. Inserting this in the expression for the errors we have
\begin{align}
\epsilon_{ b}&\approx\frac{\Delta_{{ b}} a({ b})+\Delta_{1-{ b}}(1-a({ b}))}{ D_0(\mu_{\rm c}-\mu)}\sqrt{\frac{\mu\mu_{\rm c}}{2\pi}}\e^{-(h({ b})-h_{\rm ad})}\label{eq:esalapl}\;,
\end{align}
where ${ b}=0$ or ${ b}=1$ for each of the two boundary contributions to the error. While these expressions are analytical, their evaluation is not transparent due to the complicated form of the potential. A simpler expression can be obtained by evaluating $h(1)$ and $h(0)$ by an expansion around $m_{\rm ad}$. Truncating to second order we have
\begin{align}
\epsilon_{ b}&\approx\frac{\Delta_{{ b}} a({ b})+\Delta_{1-{ b}}(1-a({ b}))}{\Delta_{0}(\mu_{\rm c}-\mu)}\sqrt{\frac{\mu\mu_{\rm c}}{2\pi}}\e^{-({ b}-m_{\rm ad})^2\mu\mu_{\rm c}/2}\quad,\nonumber
\end{align}
We now define the characteristic error
\begin{align}
\epsilon_{\rm c}=\frac{\Delta_{{ b}} a({ b})+\Delta_{1-{ b}}(1-a({ b}))}{\Delta_{0}(\mu_{\rm c}-\mu)}\sqrt{\frac{\mu\mu_{\rm c}}{2\pi}}\quad,\nonumber
\end{align}
which has a weak dependence on $\mu$. Introducing  the constant $k=({ b}-m_{\rm ad})^2/2$, we arrive at the following estimate for the adaptation error
\begin{align}
\langle\epsilon\rangle\approx\epsilon_{\rm c}\e^{-k\mu\mu_{\rm c}}\quad,\label{eq:esasi}
\end{align}
where for simplicity we consider the case of { large} signals, in which $m_{\rm ad}$ is also high and the error contribution comes dominantly from the boundary $b=1$. This expression is accurate up to order $\delta$, and corresponds to Eq.~\ref{eq:esa} in the main text. It is important to note that the higher order terms do not imply saturation, since it was shown at the beginning of this section that when $\mu\to\infty$, then $\langle A\rangle\to a_{\rm ad}$ to all orders.


%

\subsection{Noise spectrum of output}
The output generated by the system activity is given by
\begin{align}
y(t)=\frac{1}{\tau_y}\int_{-\infty}^{\infty}\Theta(t-t')\e^{-(t-t')/\tau_y}A(t')\d t'\quad,\nonumber
\end{align}
where $\Theta$ is the Heaviside step function. This corresponds to a relaxation of $y$ towards $A$ in a time scale $\tau_y$. Using the definition of Fourier transform $f(\omega)=(2\pi)^{-1/2}\int f(t)\exp(-i\omega t)\d t$,  the equation above translates to
\begin{align}
y(\omega)=\frac{\tau_y^{-1}}{\tau_y^{-1}+i\omega}A(\omega)\quad,\label{eq:linres}
\end{align}
where we have used that the time-domain convolution $(f*g)(t)$ is given by $\sqrt{2\pi}f(\omega)g(\omega)$.

To calculate the power spectrum of the output, we first note that
\begin{align}
\langle A(\omega)A(\omega')\rangle=\sqrt{2\pi}\delta(\omega+\omega')S_{A}(\omega)\quad,\nonumber
\end{align}
where $S_{A}$ is the power spectrum of the activity, which can be calculated by standard methods \cite{VanKampen,gardiner1986handbook}. From this and Eq.~\ref{eq:linres} we obtain
\begin{align}
S_{y}(\omega)=\frac{S_{A}(\omega)\tau_y^{-2}}{\tau_y^{-2}+\omega^2}\nonumber\quad,
\end{align}
which we use to calculate the power-spectrum throughout this work.

The power spectrum is related to the autocorrelation function through a Fourier transform, that is
\begin{align}
S_{y}=\frac{1}{\sqrt{2\pi}}\int_{-\infty}^{\infty} \e^{-i\omega t}C_{y}(t)\d\omega\quad;\nonumber
\end{align}
where the correlation function is defined as $C_{y}(t)=\langle y(t)y(0)\rangle$. The second moment  relates to the correlation function as
\begin{align}
\langle y^2\rangle=C_{y}(0)=\frac{1}{\sqrt{2\pi}}\int _{-\infty}^{\infty} S_{y}(\omega)\d\omega\nonumber\quad.
\end{align}
Together with  $\langle y\rangle=\langle A\rangle$, this equation allows  to calculate the variance of the output.
%
%
%

For the particular case of $\mu\to\infty$, the memory state is fixed to $m_{\rm ad}$. The activity however still fluctuates between one active and one inactive state according to Eq.~\ref{eq:twostates}. In this case, the power spectrum of the output takes a particularly simple form. To calculate it, we use that the correlation of the activity is generically given by
\begin{align}
\langle A(t) {A}(0)\rangle&=\sum_\lambda\sum_{i,j}A_i q_{\lambda,i} A_jz_{\lambda,j}p^{\rm ss}_{j}\e^{\lambda t}\nonumber
\end{align}
where ${A}_i=\{1,0\}$ is the activity observable, $p_j^{\rm ss}$ the stationary probability distribution, $\lambda$ the eigenvalues of the $W-$matrix, and ${\bf q}_\lambda$ and ${\bf z}_\lambda$ the corresponding right and left eigenvectors normalized by the condition ${\bf z}_\lambda\cdotp{\bf q}_{\lambda'}=\delta_{\lambda\lambda'}$. For the particular case of the activity we thus have \cite{VanKampen,gardiner1986handbook}
\begin{align}
\langle{A}(t) {A}(0)\rangle=\sum_\lambda  q_{\lambda,1} z_{\lambda,1}p^s_{1}\e^{\lambda t}\quad.\nonumber
\end{align}

From Eq.~\ref{eq:twostates} we have that the eigenvalues are $0$ and $-(\omega_0+\omega_1)$, the corresponding right eigenvectors $\{\omega_0,\omega_1\}/(\omega_1+\omega_0)$ and  $\{-1,1\}$, and the left eigenvectors $\{1,1\}$ and $\{-\omega_1,\omega_0\}/(\omega_1+\omega_0)$. The component $1$ of the steady state probability is simply the average activity, $p_1^s=\langle A\rangle=\omega_0/(\omega_0+\omega_1)$. The correlation function is thus given by
\begin{align}
\langle{A}(t) {A}(0)\rangle=\langle A\rangle\left(\langle A\rangle+(1-\langle A\rangle)\e^{-(\omega_0+\omega_1) t}\right)\quad,\nonumber
\end{align}
from which one can verify that $\langle A^2(0)\rangle=\langle A\rangle$ and $\langle A(\infty)A(0)\rangle=\langle A\rangle^2$. The power spectrum of the activity is then
\begin{align}
S_{A}(\omega)&=\sqrt{2\pi}\langle A\rangle^2\delta(\omega)\nonumber\\
&+\langle A\rangle(1-\langle A\rangle)\sqrt{\frac{2}{\pi}}\frac{\omega_0+\omega_1}{(\omega_0+\omega_1)^2+\omega^2}\nonumber\quad.
\end{align}
Using this expression, the relationship between the power spectrum of $y$ and $A$, and defining the activation time as $\tau_a=1/(\omega_0+\omega_1)$, we obtain the second moment of the output to be
\begin{align}
\langle y^2\rangle&=\frac{1}{\sqrt{2\pi}}\bigg(\langle A\rangle^2\sqrt{2\pi}+\frac{1}{\tau_y}\frac{\tau_a^{-1}-\tau_y^{-1}}{\tau_a^{-2}-\tau_y^{-2}}\langle A\rangle(1-\langle A\rangle)\pi\sqrt{\frac{2}{\pi}}\bigg)\nonumber\\
&=\langle A\rangle^2+\frac{\tau_a}{\tau_a+\tau_y}\langle A\rangle(1-\langle A\rangle)\nonumber
\end{align}
From this, we obtain that the variance of the output in the fully irreversible regime is given by
\begin{align}
\sigma_{ym}^2=\langle A\rangle(1-\langle A\rangle)\frac{\tau_a}{\tau_a+\tau_y}\quad,
\end{align}
which contains the intrinsic activity fluctuations, averaged by the CheY-P dynamics \cite{sartori2011noise}. This defines the saturation error noise, which can be averaged out by reducing the ratio $\tau_a/\tau_y$.

\subsection{Derivation of Eq.~\ref{eq:esf}}
As described in the main text, there are two contributions to the fluctuations of the output. The first comes from the fluctuation of $y$ around $a(m)$, we note it $\sigma^2_{ ym}$ and was calculated before in the case $\mu\to\infty$. The second comes from the fluctuations of $m$ itself, which make $a(m)$ fluctuate, and we note it $\sigma_a^2$. When there is a separation of time scales $\tau_a\ll\tau_y\ll\tau_{\rm ad}$, we have that the total output variance is the sum of these two contributions. Thus, using $\sigma_\epsilon^2=a_{\rm ad}\sigma_\epsilon^2$, we have that $\sigma_\epsilon^2=(\sigma_a^2+\sigma_{ym}^2)/a_{\rm ad}$. For finite values of  $\mu$ the memory fluctuations dominate and $\sigma^2_a$ is the most relevant contribution, however as $\mu\to\infty$ the memory gets frozen and $\sigma_{ym}^2$ dominates. It is thus crucial to calculate the dependence of $\sigma_a^2$ in $\mu$.  The main steps on how to calculate $\sigma_a^2$ are given in the main text.

\section{Dissipated work}
To quantify how far from equilibrium the system is, we use the dissipated work $\dot{W}$. At the steady state, and having set the thermal unit to one, we have
\begin{align}
\dot{W}=\int_0^1\left[\frac{J_0^2}{ D_0p_0}+\frac{J_1^2}{ D_1p_1}+(p_0\omega_0-p_1\omega_1)\log\left(\frac{p_0\omega_0}{p_1\omega_1}\right)\right]\d m\;.\nonumber
\end{align}
We can estimate this quantity in the highly irreversible limit $\mu\gg\mu_{\rm c}$ to zeroth order in $\delta$. In that regime, we have
\begin{align}
\dot{W}&\approx\int_0^1\left[\frac{(\mu D_0p_0)^2}{ D_0p_0}+\frac{(\mu D_1p_1)^2}{ D_1p_1}\right]\d m\nonumber\\
&=\mu^2\int_0^1\left[ D_0p_0+ D_1p_1\right]\d m\nonumber\\
&=\frac{|\mu|}{\tau_{\rm ad}}(\bar{D}_0(1-a_{\rm ad})+\bar{D}_1a_{\rm ad})\;.\nonumber
\end{align}
For the case considered in the main text in which $ D_0= D_1$, we have that $\dot{W}\approx|\mu|\bar{D}_0/\tau_{\rm ad}$.

\section{Parameters}
Unless otherwise specified, the parameters used in the continuum limit are:  $e=8$,  $m_{\rm r}=1/2$,  $d=1/e$, $S_{\rm r}=0$, $S=2.5$, $ D_0= D_1$, $\tau_a=10^{-3}s$, $\tau_y=10^{-1}s$, $\tau_{\rm ad}=10s$. The parameters for the discrete model are used taking these as references and using the appropriate $N$ rescaling. For $N=5$ as in {\it E. Coli} chemotaxis the resulting parameters are those in \cite{Tu30092008}.

\end{document}